\documentclass[nofootinbib]{revtex4}
\usepackage{graphicx}
\usepackage{fancyhdr}
\usepackage{amsmath,amssymb}
\usepackage{mathrsfs}
\usepackage{dsfont}

\newenvironment{Eqnarray}%
         {\arraycolsep 0.14em\begin{eqnarray}}{\end{eqnarray}}
\def\beqa{\begin{Eqnarray}}
\def\eeqa{\end{Eqnarray}}
\def\beq{\begin{equation}}
\def\eeq{\end{equation}}
\def\vev#1{\langle #1 \rangle}
\def\nn{\nonumber}
\def\half{\tfrac{1}{2}}
\renewcommand\Re{{\rm Re}}
\renewcommand\Im{{\rm Im}}

\def\eq#1{eq.~(\ref{#1})}

\def\eqst#1#2{eqs.~(\ref{#1})--(\ref{#2})}
\def\ls#1{\ifmath{_{\lower1.5pt\hbox{$\scriptstyle #1$}}}}

\def\ur{U_R}
\def\dr{D_R}

\def\anti{\overline}
\def\ifmath#1{\relax\ifmmode #1\else $#1$\fi}
\def\lsub#1{\ifmath{_{\lower1.5pt\hbox{$\scriptstyle #1$}}}}
\def\phm{\phantom{-}}

\def\cosbma{\cos(\beta-\alpha)}
\def\cosbmasq{\cos^2(\beta-\alpha)}
\def\sinbma{\sin(\beta-\alpha)}

\def\llsup#1{^{\lower 2pt\hbox{$\scriptstyle#1$}}}
\def\lsim{\mathrel{\raise.3ex\hbox{$<$\kern-.75em\lower1ex\hbox{$\sim$}}}}
\def\gsim{\mathrel{\raise.3ex\hbox{$>$\kern-.75em\lower1ex\hbox{$\sim$}}}}

\usepackage{color}

\begin{document}

\title{The Higgs data and the Decoupling Limit}

%

\author{Howard E.~Haber}
\affiliation{Santa Cruz Institute for Particle Physics \\ University of California, Santa Cruz, CA 95064 USA}

\begin{abstract}
The Higgs data analyzed by the ATLAS and CMS Collaborations suggest
that the scalar state discovered in 2012 is a Standard Model
(SM)--like Higgs boson.  Nevertheless, there is still significant room
for Higgs physics beyond the Standard Model.  Many approaches to
electroweak symmetry breaking possess a decoupling limit in which the
properties of the lightest CP-even Higgs scalar approach those of the
SM Higgs boson.  In some cases, an apparent SM-like Higgs signal can
also arise in a regime that may not be governed by the decoupling
limit.  One such scenario can be realized if the observed Higgs signal
is a result of two unresolved nearly-mass-degenerate scalar states.
The general two-Higgs doublet model provides a useful framework for
studying the decoupling limit and possible departures from SM-like
Higgs behavior.  The implications for current and future Higgs data
are briefly considered.

\end{abstract}

\maketitle

\thispagestyle{fancy}


\section{Introduction}

Since the initial announcement of a newly discovered boson on July 4, 2012~\cite{discovery} with mass $m\simeq 125$~GeV, data collected by the ATLAS and CMS Collaborations have revealed a phenomenological profile for the Higgs boson that closely resembles that of the Higgs boson of the Standard Model (SM).  The present Higgs data are still statistically limited.  Moreover, the LHC can only provide measurements of cross section times branching ratio, since the absolute width of the Higgs boson (which for a SM Higgs mass of 125 GeV is about 4~MeV~\cite{PDG}) cannot be directly measured and is difficult to determine at the LHC by indirect means in a model-independent way.  Nevertheless, by making some weak assumptions (e.g.~no appreciable Higgs decays into light fermion pairs and invisible modes), one can extract values for Higgs couplings from the LHC Higgs data~\cite{PDG}.

\begin{figure}[t!]
\begin{center}
\includegraphics*[width=0.9\textwidth]{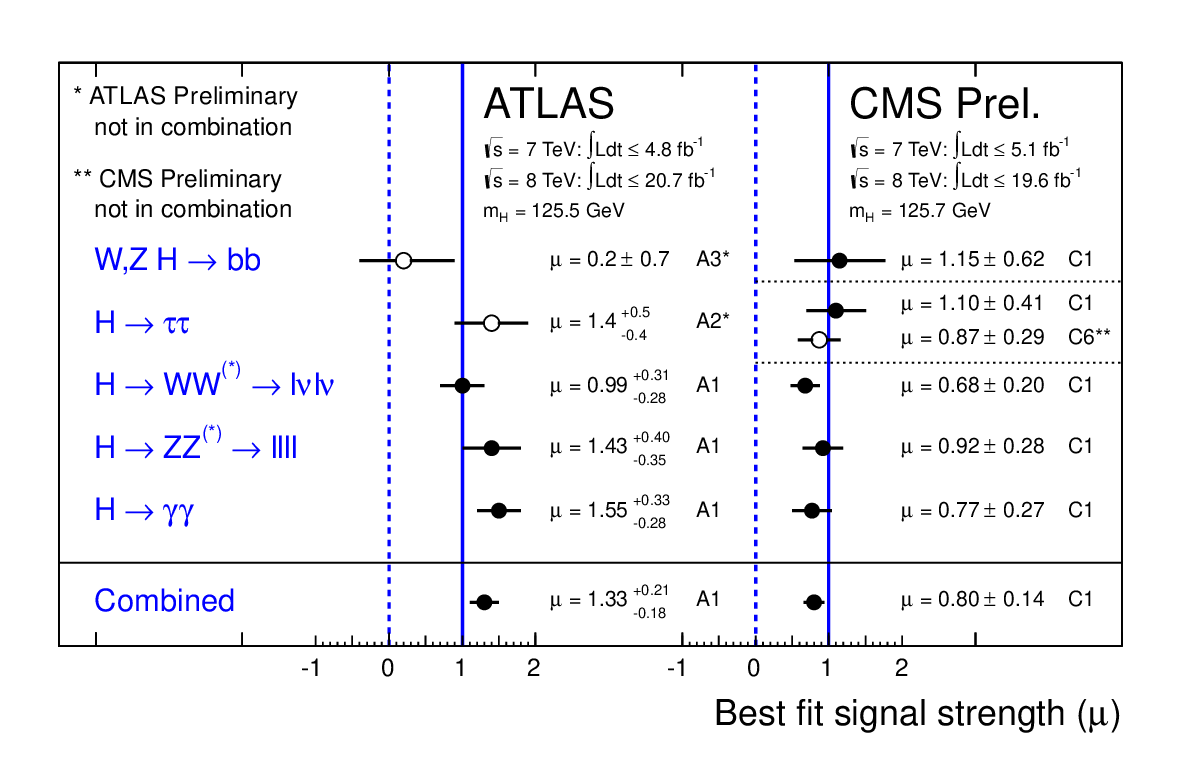}
\end{center}
\caption{\label{signal_strengths}
The signal strengths $\mu$ measured
by the ATLAS experiment from Refs. A1~\cite{A1},
A2~\cite{A2} and A3~\cite{A3}, and CMS experiment
from Refs. C1~\cite{C1,C2} and C6~\cite{C6} in the five
principal channels and their combination. It should
be noted that the ATLAS combination only includes the bosonic
$\gamma\gamma$, $ZZ$ and $WW$ channels.
Taken from Ref.~\cite{PDG}.
}
\end{figure}

A summary of Higgs signal strengths measured by the ATLAS and CMS experiments is shown in Fig.~\ref{signal_strengths}~\cite{A1,A2,A3,C1,C2,C6}.
The Higgs data set shows some interesting fluctuations relative to SM expectations.  An enhanced $\gamma\gamma$ signal in the ATLAS data~\cite{A1,A4} is perhaps the most notable departure from the SM, although a similar enhancement initially reported by the CMS collaboration has since
disappeared~\cite{C3}.   However, even the largest deviation does not reach the $3\sigma$ level.  Global fits of the full Higgs data set yield results that are consistent with Standard Model predictions~\cite{PDG}.  Thus, it is probably fair to conclude that a SM-like Higgs boson has been discovered, and any deviations of its properties from Standard Model expectations are likely to be small.

In this talk, I will discuss the concept of the decoupling limit of the Higgs sector~\cite{Haber:1989xc}, which yields one neutral Higgs state whose properties are close to those of the SM-Higgs boson.  General features of the decoupling limit are presented in Section II.  The two-Higgs doublet model (2HDM) provides an explicit framework in which to study the decoupling limit~\cite{Gunion:2002zf}.  In Section III, two forms of the decoupling limit are exhibited:
large-mass decoupling in which all but one of the Higgs boson masses of the 2HDM are very heavy and weak-coupling decoupling in which a SM-like Higgs boson arises in a limit where one scalar self-coupling in the Higgs basis~\cite{higgsbasis}
approaches zero.  In order to accommodate naturally small tree-level Higgs-mediated flavor changing neutral currents, it is standard practice to adopt a special form for the Higgs--fermion Yukawa couplings~\cite{GWP}, as reviewed in Section IV.  The corresponding decoupling limits are special cases of the ones obtained in the most general 2HDM.  In Section V, it is shown that a SM-like Higgs boson can be consistent with the existence of two nearly-mass-degenerate neutral Higgs states~\cite{degenrefs1,Ferreira:2012nv}.  The implications of the present and future LHC data for such scenarios are considered.   Finally, conclusions are presented in Section VI.

\section{The Decoupling Limit of the Higgs Sector}

The Higgs boson serves as a window to physics beyond the SM only if one can experimentally establish
deviations of Higgs couplings from their SM values, or discover new scalar degrees of freedom beyond the
SM--like Higgs boson.  The prospects to achieve this are challenging in general due to the
decoupling limit~\cite{Haber:1989xc}.
In extended Higgs models (as well as in some alternative models of electroweak symmetry breaking), most of the
parameter space typically yields a neutral CP Higgs boson with SM--like tree level couplings and additional
scalar states that are somewhat heavier in mass (of order $\Lambda_H$), with small mass splittings of order
$(m_Z/\Lambda_H)m_Z$.  Below the scale $\Lambda_H$, the effective Higgs theory coincides with the Higgs
sector of the SM.

In this discussion, two energy scales must be distinguished.  The first is $\Lambda_H$, introduced above,
which characterizes the mass scale of the (presumed) heavier non-minimal Higgs bosons.  The second is
$\Lambda_{\rm NP}$, which characterizes the scale of new physics beyond the Higgs-extended SM.
The departure from the decoupling limit can receive contributions from both the Heavy Higgs states
via tree-level mixing and from one-loop radiative corrections in which particles associated with the
new physics beyond the SM can enter.  If deviations from SM Higgs
couplings are confirmed, the separation of these two effects will be an important and challenging task.

Small deviations from SM--like Higgs behavior can be a consequence of the decoupling of heavy states, which yields corrections to Higgs couplings of $\mathcal{O}(m_Z^2/\Lambda_H^2)$.
We denote this phenomenon by \textit{large-mass decoupling}.
But there is an alternative mechanism which can contribute small corrections to SM--like Higgs behavior
that arises in the weak coupling limit of new physics phenomena.  In this case, there is no requirement
that all new states associated with the new physics phenomena must be heavy.  An example
of such a scenario is the Higgs portal~\cite{portal}, in which the SM Higgs bosons couples to a hidden sector
consisting of new states that are neutral with respect to the SM gauge group.  For example,
if $H$ is the SM Higgs boson and $\phi$ is a neutral real scalar, then an interaction term of the form
$$
\mathscr{L}_{\rm int}=\lambda H^\dagger H \phi^2\,,
$$
can modify the decay properties of the SM Higgs boson (by allowing $H\to\phi\phi)$ if $m_H>2m\phi$.
The deviation of the total Higgs width from its SM value vanishes in the limit of $\lambda\to 0$.
Thus the weak coupling limit provides an alternative mechanism for decoupling, which we henceforth denote by \textit{weak-coupling decoupling}.
Similar
modifications to Higgs properties can also be achieved by coupling the SM Higgs boson to new particles
that are charged under the SM gauge group.  As an example, the lightest electrically neutral particle of
the new particle sector can also provide a possible invisible decay channel for the SM Higgs.

The two-Higgs doublet model~\cite{thdm,Haber:1978jt,Donoghue:1978cj,thdmreview} (2HDM) provides a compelling laboratory for studying the phenomenology of an extended Higgs
sector and possible departures from the decoupling limit.  Such models are often motivated by the
minimal supersymmetric extension of the Standard Model (MSSM), which requires a second Higgs doublet in order to cancel anomalies that arise from the higgsino
partners~\cite{Fayet:1974pd}.
The MSSM exhibits examples of
the various sources described above that can yield departures from the decoupling limit.  Higgs mixing
at tree-level can alter SM Higgs couplings.  One-loop corrections due to the exchange of heavy
supersymmetric particles can also yield corrections to SM Higgs behavior~\cite{Haber:2007dj} .  Finally, if the lightest
neutralino is sufficiently light, it can provide for an invisible Higgs decay channel thereby altering
the Higgs total width and branching ratios.


\section{The general Two-Higgs doublet model}

The most general version of the 2HDM, which contains all possible renormalizable terms (mass terms
and interactions) allowed by the electroweak gauge invariance,
is not phenomenologically viable due to the presence of Higgs-fermion Yukawa interaction terms that
lead to tree-level Higgs-mediated flavor changing neutral currents (FCNCs).  Such effects are absent
in the MSSM Higgs sector due to supersymmetry (SUSY), which constrains the form of the
Higgs-fermion Yukawa interactions.  In non-supersymmetric versions
of the 2HDM, one can also naturally avoid FCNCs by imposing certain simple discrete symmetries on
the Higgs Lagrangian.  These symmetries reduce the parameter freedom of the 2HDM
and automatically eliminate the dangerous FCNC interactions.  However, the symmetries used to
constrain the Higgs Lagrangian (either SUSY or the discrete symmetries) are typically broken
(either via soft explicit breaking and/or by the vacuum).
In these cases, the Higgs-mediated FCNC interactions 
are generated at the loop level, while remaining absent at tree-level,
thereby satisfying the phenomenological constraints.

If the energy scale associated with the symmetry-breaking is above the mass scale of
the heaviest Higgs bosons of the 2HDM, then one can integrate out the heavy degrees of freedom of the new physics.
The low-energy effective field theory describing
the Higgs physics is an unconstrained 2HDM, which
contains all possible gauge-invariant terms of dimension four or less~\cite{Haber:2007dj} .  In particular,
terms that were previously absent due to symmetry reasons are now present; the size of
the corresponding coefficients can be determined by matching the low-energy effective theory to the complete theory above
the symmetry-breaking scale.  Not surprisingly, the size of these coefficients possess one-loop
suppression factors, although in some special cases, enhancement factors (such as $\tan\beta$
in the MSSM~\cite{Carena:2002es}) can render them phenomenologically relevant.

We are therefore motivated to examine the
general 2HDM without imposing additional
constraints on the Higgs Lagrangian.
Consider the complex scalar doublet, hypercharge-one fields,
$\Phi_1$ and $\Phi_2$ of the 2HDM in a generic basis,
where the vacuum expectation value (vev) of the neutral component of
the scalar doublet is denoted by
$\langle\Phi^0_i\rangle=v_i/\sqrt{2}$ (for $i=1,2$), and
$v^2\equiv |v_1|^2+|v_2|^2=(246~{\rm GeV})^2$.
It is convenient to
define new Higgs doublet fields~\cite{Davidson:2005cw,Haber:2006ue},
$$H_1=\begin{pmatrix}H_1^+\\ H_1^0\end{pmatrix}\equiv \frac{v_1^* \Phi_1+v_2^*\Phi_2}{v}\,,
\qquad\quad H_2=\begin{pmatrix} H_2^+\\ H_2^0\end{pmatrix}\equiv\frac{-v_2 \Phi_1+v_1\Phi_2}{v}
 \,.
$$
such that $\vev{H_1^0}=v/\sqrt{2}$ and $\vev{H_2^0}=0$, where
$v=246$~GeV is real and positive.
This is the \textit{Higgs basis}~\cite{higgsbasis}, which is uniquely defined
up to an overall rephasing, $H_2\to e^{i\chi} H_2$.
In the Higgs basis, the scalar potential is
given by:
\beqa \mathcal{V}&=& Y_1 H_1^\dagger H_1+ Y_2 H_2^\dagger H_2 +[Y_3
H_1^\dagger H_2+{\rm h.c.}]
+\half Z_1(H_1^\dagger H_1)^2\nn\\
&&\quad
+\half Z_2(H_2^\dagger H_2)^2
+Z_3(H_1^\dagger H_1)(H_2^\dagger H_2)
+Z_4( H_1^\dagger H_2)(H_2^\dagger H_1)\nn\\
&&\quad +\left\{\half Z_5 (H_1^\dagger H_2)^2 +\big[Z_6 (H_1^\dagger
H_1) +Z_7 (H_2^\dagger H_2)\big] H_1^\dagger H_2+{\rm
h.c.}\right\}\,,\nn
\eeqa
where $Y_1$, $Y_2$ and $Z_1,\ldots,Z_4$ are real and uniquely defined,
whereas $Y_3$, $Z_5$, $Z_6$ and $Z_7$ are complex and transform under
the rephasing of $H_2$,
\vspace{-0.1in}
$$ [Y_3, Z_6, Z_7]\to e^{-i\chi}[Y_3, Z_6, Z_7] \quad{\rm and}\quad
Z_5\to  e^{-2i\chi} Z_5\,.$$
After minimizing the scalar potential, $Y_1=-\tfrac{1}{2}Z_1 v^2$
and $Y_3=-\tfrac{1}{2} Z_6 v^2$.  This leaves 11 free parameters:
1 vev, 8 real parameters, $Y_2$, $Z_{1,2,3,4}$, $|Z_{5,6,7}|$, and two relative phases.

If $\Phi_1$ and $\Phi_2$ are indistinguishable fields, then
observables can only depend on combinations of Higgs basis
parameters that are independent of $\chi$.  Symmetries, such as
discrete symmetries or supersymmetry, can distinguish between
 $\Phi_1$ and $\Phi_2$, which then singles out a specific basis for
the Higgs fields, and can yield additional observables such as
$\tan\beta$ in the MSSM.  However, as noted at the end of Section II,
such symmetries are typically broken, so that below the symmetry-breaking
scale, the effective 2HDM is generic, and all possible scalar potential
terms can appear.

In the general 2HDM,
the physical charged Higgs boson is the charged component of the Higgs-basis doublet $H_2$, and its mass
is given by $m_{H^\pm}^2=Y_{2}+\half Z_3 v^2$.
The three physical neutral Higgs boson mass-eigenstates
are determined by diagonalizing a $3\times 3$ real symmetric squared-mass
matrix that is defined in the Higgs basis,
$$
\mathcal{M}^2=v^2\left( \begin{array}{ccc}
Z_1&\,\, \Re(Z_6) &\,\, -\Im(Z_6)\\
\Re(Z_6)  &\,\, \half Z_{345}+Y_2/v^2 & \,\,
- \half \Im(Z_5)\\ -\Im(Z_6) &\,\, - \half \Im(Z_5) &\,\,
 \half Z_{345}-\Re(Z_5)+Y_2/v^2\end{array}\right),
$$
where $Z_{345}\equiv Z_3+Z_4+\Re(Z_5)$.  The diagonalizing matrix is a $3\times 3$
real orthogonal matrix that depends on three angles,
$\theta_{12}$, $\theta_{13}$ and~$\theta_{23}$, as defined in Ref.~\cite{Haber:2006ue}.
The corresponding neutral Higgs masses will be denoted by
$m_1$, $m_2$ and $m_3$.  By convention, we take $m_1\leq m_{2,3}$.
Under the rephasing $H_2\to e^{i\chi}H_2$,
$$
\theta_{12}\,,\, \theta_{13}~{\hbox{\text{are invariant, \,and}}}\,\,
\theta_{23}\to  \theta_{23}-\chi\,.
$$
This procedure defines a second physical basis of the general 2HDM---namely, the
mass-eigenstate basis for the neutral Higgs bosons.

The large-mass decoupling limit corresponds to $Y_2\gg v$.  In this limit, the the Higgs basis and
the mass-eigenstate basis coincide.  Namely, $H_1$ becomes the SM Higgs doublet, and
$H_2$ becomes a massive scalar doublet that decouples.  Thus, in the approach to the
decoupling limit, we can identify a number of important parameters that are small by
virtue of the fact that $v/Y_2\ll 1$.  If we identify $h_1$ as the SM-like Higgs boson with $m_1\simeq 125$~GeV, then~\cite{Haber:2006ue,prep}
 \beqa
s_{12}\equiv\sin\theta_{12}&\simeq &\frac{\Re(Z_6 e^{-i\theta_{23}})v^2}{m_2^2-m_1^2}\ll 1\,,\label{d1} \\
s_{13}\equiv\sin\theta_{13}&\simeq& -\frac{\Im(Z_6 e^{-i\theta_{23}})v^2}{m_3^2-m_1^2}\ll 1\,,\label{d2}\\
\Im(Z_5 e^{-2i\theta_{23}})&\simeq& \frac{2(m_2^2-m_1^2) s_{12}s_{13}}{v^2}
\simeq-\frac{\Im(Z_6^2 e^{-2i\theta_{23}})v^2}{m_3^2-m_1^2}\ll 1\,,\label{d3} \\
m_2^2-m_3^2 &\simeq& \Re(Z_5 e^{-2i\theta_{23}})v^2\,,\label{d4}
\eeqa
where $m_1^2\simeq Z_1 v^2\ll m_2^2, m_3^2$.

In fact, both large-mass decoupling and weak-coupling decoupling\footnote{The weak-coupling decoupling limit of the 2HDM
was first examined in detail in Ref.~\cite{Gunion:2002zf}.  This limit was later called the alignment limit in Ref.~\cite{Craig:2013hca}.
Recent phenomenological studies of this limit in the CP-conserving 2HDM can be found in Refs.~\cite{Craig:2013hca} and \cite{Carena:2013ooa}.}
are on display in \eqst{d1}{d3}.
Indeed, in the limit of $Z_6\to 0$, the tree-level couplings of $h_1$ are precisely
those of the SM Higgs boson since $s_{12}=s_{13}=\Im(Z_5 e^{-2i\theta_{23}})=0$,
independently of values of $m_1$, $m_2$ and $m_3$.   Note that in the case of weak-coupling decoupling, it is possible to have $m_1$, $m_2$ and $m_3$ all of the same order.

In the analysis above, we have assumed that the SM-like Higgs boson is the lightest neutral Higgs state.  Suppose that $h_2$ is identified as the SM-like Higgs boson.  Then~\cite{prep,Asner:2013psa},
\beqa
c_{12}\equiv\cos\theta_{12}&\simeq &\frac{\Re(Z_6 e^{-i\theta_{23}})v^2}{m_2^2-m_1^2}\ll 1\,,\label{d5} \\
s_{13}\equiv\sin\theta_{13}&\simeq& -\frac{\Im(Z_6 e^{-i\theta_{23}})v^2}{m_3^2-m_2^2}\ll 1\,,\label{d6}\\
\Im(Z_5 e^{-2i\theta_{23}})&\simeq& \frac{2(m_2^2-m_1^2) c_{12}s_{13}}{v^2}
\simeq-\frac{\Im(Z_6^2 e^{-2i\theta_{23}})v^2}{m_3^2-m_2^2}\ll 1\,,\label{d7} \\
m_1^2-m_3^2 &\simeq& \Re(Z_5 e^{-2i\theta_{23}})v^2\,.\label{d8}
\eeqa
In this case, the large-mass decoupling limit does not exist, since we are identifying $m_2\simeq 125$~GeV.   However, \eqst{d5}{d8} can be achieved in the weak-coupling decoupling limit when $|Z_6|\ll 1$.  In this case, all three neutral Higgs masses must be of order 125 GeV (or below).  This case allows for the possibility that
of a new decay channel for the SM-like Higgs boson via $h_2\to h_1 h_1$ if $m_{h_1}<\half m_{h_2}$.
If this new decay channel is present, then all branching ratios of the SM-like Higgs boson will deviate from
SM expectations (despite the fact that the partial widths into the standard decay channels are unmodified from their SM predictions).

We next turn to the Higgs-fermion Yukawa couplings.  We focus
on the interaction of the Higgs bosons with three generations of quarks, since the
corresponding interactions with leptons are easily obtained from the latter by
the appropriate substitutions.   One starts out initially with a Lagrangian
expressed in terms of the scalar doublet fields $\Phi_i$ ($i=1,2$) and the
interaction--eigenstate quark fields.  After electroweak symmetry breaking,
one can transform the scalar doublets into the Higgs basis fields $H_1$ and $H_2$.
At the same time,
one can identify the $3\times 3$ quark mass matrices.  By redefining the left
and right-handed quark fields appropriately, the quark
mass matrices are transformed into diagonal form, where the diagonal elements are real
and non-negative.  The resulting Higgs--quark Yukawa couplings are given by~\cite{Haber:2006ue,Haber:2010bw}
\beqa
-\mathcal{L}_{\rm Y}&=&\overline U_L (\kappa^U H_1^{0\,\dagger}
+\rho^U H_2^{0\,\dagger})\ur
-\anti D_L K^\dagger(\kappa^U H_1^{-}+\rho^U H_2^{-})\ur \nonumber \\
&& +\anti U_L K (\kappa^{D\,\dagger}H_1^++\rho^{D\,\dagger}H_2^+)\dr
+\anti D_L (\kappa^{D\,\dagger}H_1^0+\rho^{D\,\dagger}H_2^0)\dr+{\rm h.c.},
\label{lyuk}
\eeqa
where $U=(u,c,t)$ and $D=(d,s,b)$ are the mass-eigenstate quark fields, $K$ is the
Cabibbo-Kobayashi-Maskawa (CKM)  
mixing matrix and $\kappa$ and $\rho$ are $3\times 3$
Yukawa coupling matrices.  Note that $Q_{R,L}\equiv P_{R,L}Q$,
where $Q=U$ or~$D$ and $P_{R,L}\equiv\half(1\pm\gamma\lsub{5})$ are
the right and left handed projection operators, respectively.

By setting $H_1^0=v$ and $H_2^0=0$, one can relate
$\kappa^U$ and $\kappa^D$ to the diagonal
quark mass matrices $M_U$ and $M_D$,
respectively,
\beq \label{mumd}
M_U=v\kappa^U={\rm diag}(m_u\,,\,m_c\,,\,m_t)\,,\qquad
M_D=v\kappa^{D\,\dagger}={\rm
diag}(m_d\,,\,m_s\,,\,m_b) \,.
\eeq
However, the complex matrices $\rho^Q$ ($Q=U,D$) are unconstrained.   Moreover,
\beq \label{rhoq}
\rho^Q\to e^{-i\chi}\rho^Q\,,
\eeq
under the rephasing $H_2\to e^{i\chi}H_2$.

In general the $\rho^Q$ are complex non-diagonal matrices.  As a result, the most general
2HDM exhibits tree-level Higgs-mediated FCNCs and new sources of CP-violation in the interactions of the neutral Higgs bosons.  In the decoupling limit, CP-violating and tree-level FCNCs mediated by the SM-like Higgs boson ($h_1$) are suppressed by factors of  $s_{12}$ and $s_{13}$.  In contrast, the interactions of the other neutral Higgs bosons ($h_2$ and $h_3$) in the decoupling limit can exhibit both CP-violating and flavor non-diagonal couplings proportional to the $\rho^Q$.   Note that in the large-mass decoupling limit, all Higgs-mediated FCNCs are suppressed by factors of $\mathcal{O}(v^2/m_{2,3}^2)$, whereas in the weak-coupling decoupling limit where $|Z_6|\ll 1$ but $m_{2,3}\sim\mathcal{O}(m_1)$, FCNCs mediated by $h_2$ and $h_3$
are generically unsuppressed.

There are four possible strategies for avoiding tree-level Higgs-mediated FCNCs in the 2HDM.
First, one can arbitrarily declare the $\rho^Q$ to be flavor diagonal matrices.  Such conditions are not renormalization group stable and thus must be regarded as unnaturally fine-tuned~\cite{Ferreira:2010xe}.  Second, one can impose an appropriate discrete symmetry or supersymmetry, which can yield the so-called Type-I~\cite{Haber:1978jt,Hall:1981bc} or Type-II~\cite{Donoghue:1978cj,Hall:1981bc} Higgs-quark interactions.\footnote{In the Type-I
and II 2HDMs, the Higgs couplings to charged leptons follow the same pattern as the Higgs couplings to down-type quarks.  There are
two additional coupling patterns, called Types III and IV in Ref.~\cite{Barger:1989fj} and called Types Y and X in Ref.~\cite{Aoki:2009ha}, respectively, in which the Higgs couplings to down-type quarks and charged leptons
do not match.  These latter two coupling patterns will not be considered further in this talk.}
Such symmetries select out a special basis in which the symmetries are manifest.  The relative orientation of this basis relative to the Higgs basis defines the ratio of neutral Higgs vacuum expectation values, $\tan\beta$.   Indeed, in such a framework, $\rho^Q \propto M_Q$ is automatically diagonal.

In the aligned 2HDM~\cite{Pich:2009sp}, one imposes the condition $\rho^Q=\alpha^Q \kappa^Q$, where the $\alpha^Q$ are complex scalar parameters.   Such a condition generalizes the Type-I and II 2HDMs.  However, the more general alignment condition is not renormalization group stable unless
the resulting Higgs-quark interactions are of Type I or II~\cite{Ferreira:2010xe}.  This is not surprising, since there is no symmetry that governs the more general alignment condition.  Thus, the aligned 2HDM must be considered to be unnaturally fine-tuned.  One could imagine new physics beyond the 2HDM that imposes the alignment condition at some higher energy scale.  Renormalization group running would then generate non-diagonal $\rho^Q$ at the electroweak scale.  However, in this case the departures from flavor diagonal could be small enough to be phenomenologically acceptable~\cite{Braeuninger:2010td}.

Finally, as noted above, Higgs-mediated FCNCs are suppressed in the heavy-mass decoupling limit
by factors of $\mathcal{O}(v^2/m_{2,3}^2)$, where $m_{2,3}$ could be in the multi-TeV range.  A detailed phenomenological study is required to ascertain the precise limits on the heavy scalar masses.

\section{Special forms for the Higgs-quark Yukawa interactions}

Since the Type-I and II Higgs-quark interactions provide a natural solution for suppressing Higgs-mediated FCNCs, we briefly discuss the structure of the corresponding 2HDMs.

The scalar potential exhibits a $\mathbb{Z}_2$ symmetry that is at most softly broken,
\beqa
\mathcal{V}&=& m_{11}^2 \Phi_1^\dagger \Phi_1+ m_{22}^2 \Phi_2^\dagger \Phi_2 -[m_{12}^2
\Phi_1^\dagger \Phi_2+{\rm h.c.}]
+\half \lambda_1(\Phi_1^\dagger \Phi_1)^2+\half \lambda_2(\Phi_2^\dagger \Phi_2)^2
+\lambda_3(\Phi_1^\dagger \Phi_1)(\Phi_2^\dagger \Phi_2)
\nn\\
&&\quad
+\lambda_4( \Phi_1^\dagger \Phi_2)(\Phi_2^\dagger \Phi_1)
 +\left\{\half \lambda_5 (\Phi_1^\dagger \Phi_2)^2 +{\rm
h.c.}\right\}\,,\label{genpot}
\eeqa
where $m_{12}^2$ and $\lambda_5$ are assumed to be real.
After minimizing the scalar potential,\footnote{For simplicity, we assume that $\lambda_5<|m_{12}^2/(v_1 v_2)|$.
In this case, the minimum of the scalar potential does not break CP~\cite{Gunion:2002zf} and one can rephase the Higgs fields
such that $v_1$ and $v_2$ are both real and non-negative.}
$\langle\Phi_a^0\rangle=v_a/\sqrt{2}$ (for $a=1,2)$
where $v^2\equiv v_1^2+v_2^2=4m_W^2/g^2=(246~{\rm GeV})^2$ and $\tan\beta\equiv v_2/v_1$ is a
free parameter of the model.  The Higgs spectrum consists of a charged Higgs pair $H^\pm$ ad
three neutral Higgs states: two CP-even Higgs states, $h^0$ and $H^0$ (with $m_{h^0}\leq m_{H^0}$) and a CP-odd Higgs state $A^0$.   The two CP-even mass eigenstates are determined by
diagonalizing a $2\times 2$ squared-mass matrix; the corresponding CP-even Higgs mixing
angle is denoted by $\alpha$.   By convention, we may choose
\beq
0\leq\beta\leq\half\pi\,,\qquad\quad 0\leq\beta-\alpha<\pi\,.
\eeq
Thus the scalar sector is governed eight real parameters:
$v$, $\alpha$, $\beta$, four physical Higgs masses and one additional parameter, typically
taken to be either $M^2\equiv 2m_{12}^2/\sin 2\beta$ or $\lambda_5=(M^2-m_A^2)/v^2$.

The most general Yukawa Lagrangian, in terms of the quark mass-eigenstate fields and a generic basis of scalar doublet fields, is given by
\beq \label{higgsql}
-\mathcal{L}_{\rm Y}=\overline U_L \Phi_{a}^{0\,*}{{h^U_a}} \ur -\anti
D_L K^\dagger\Phi_{a}^- {{h^U_a}}\ur
+\overline U_L K\Phi_a^+{{h^{D\,\dagger}_{a}}} \dr
+\overline D_L\Phi_a^0 {{h^{D\,\dagger}_{a}}}\dr\,,
\eeq
where there is an implicit sum over $a=1,2$ and $K$ is the CKM matrix.  We introduce a discrete
$\mathbb{Z}_2$ symmetry under which $\Phi_1\to +\Phi_1$ and $\Phi_2\to -\Phi_2$, which is broken softly by the term proportional to $m_{12}^2$ in \eq{genpot}.  The quark fields transform under the discrete symmetry as  $Q_L\to +Q_L$ and
$U_R\to -U_R$.   Two possible choices for the transformation of $D_R$ yield the
Type-I and Type-II 2HDMs, corresponding to $D_R\to -D_R$
or $D_R\to +D_R$, respectively.  Imposing this symmetry on the Higgs-quark interactions implies that two of the four matrix Yukawa couplings, $h_{1,2}^U$ and $h_{1,2}^D$, vanish.  In particular,
\vspace{0.1in}

Type-I 2HDM ( $h_1^U=h_1^D=0$), which yields Higgs-quark couplings:
\begin{table}[h!]
\begin{center}
\begin{tabular}{|c|ccccccc|}
\hline
 Type-I couplings & & $h^0$  & \hspace{15pt} & $A^0$ & \hspace{15pt} & $H^0$ & \\
\hline
Up-type quarks  & &
$\phm\cos{\alpha}/\sin{{\beta}}$   & &
$\phm\cot{\beta}$  & &
$\phm\sin{\alpha}/\sin{{\beta}}$   &\\
Down-type quarks and charged leptons & &
$\phm\cos{\alpha}/\sin{{\beta}}$   & &
$- \cot{\beta}$  & &
$\phm\sin{\alpha}/\sin{{\beta}}$  &  \\
\hline
\end{tabular}
\end{center}
\end{table}

Type-II 2HDM ($h_1^U=h_2^D=0$), which yields Higgs-quark couplings:
\begin{table}[h!]
\begin{center}
\begin{tabular}{|c|ccccccc|ccccccc|}
\hline
Type-II couplings & & $h^0$  & \hspace{15pt} & $A^0$ & \hspace{15pt} & $H^0$ & \\
\hline
Up-type quarks  & &
$\phm\cos{\alpha}/\sin{{\beta}}$   & &
$\phm\cot{\beta}$ & &
$\phm\sin{\alpha}/\sin{{\beta}}$ &\\
Down-type quarks and charged leptons & &
$- \sin{\alpha}/\cos{{\beta}}$   & &
$\phm\tan{\beta}$ & &
$\phm\cos{\alpha}/\cos{{\beta}}$ & \\
\hline
\end{tabular}
\end{center}
\end{table}

Since the models under consideration exhibit a CP-conserving scalar potential and vacuum,
one can find a Higgs basis in which $Y_3$ and $Z_{5,6,7}$ are all real.  Within this real basis,
one is still free to redefine $H_2\to -H_2$, which changes the signs of $Y_3$, $Z_6$ and $Z_7$.
In particular, the sign of $Z_6$, denoted by
\beq
\varepsilon_6\equiv {\rm sgn}~Z_6\,,
\eeq
will characterize the two-fold ambiguity of the real Higgs basis.  No physical parameter can depend on
$\varepsilon_6$.

We can relate the notation of the CP-conserving model to the general 2HDM of Section III by identifying~\cite{Haber:2006ue}
$h_1=h^0$, $h_2=-\varepsilon_6 H^0$, $h_3=\varepsilon_6 A^0$, $e^{-i\theta_{23}}=\varepsilon_6$,
$\theta_{13}=0$ and
\beq
c_{12}\equiv\cos\theta_{12}=\sin(\beta-\alpha)\,,
\qquad\quad
s_{12}\equiv\sin\theta_{12}=-\varepsilon_6\cos(\beta-\alpha)\,.
\eeq
If $h^0$ is the SM-like Higgs boson, then \eq{d1} yields the condition for the decoupling limit,
\beqa
\cos(\beta-\alpha)&\simeq &-\frac{Z_6 v^2}{m_H^2-m_h^2}\ll1\,,\\
m_H^2-m_A^2\ & \simeq & Z_5 v^2\,,
\eeqa
which can be achieved for large-mass decoupling ($m_{H,A}\gg m_h$) and/or for
weak-coupling decoupling ($|Z_6|\ll 1$).   Note that the sign of $\cos(\beta-\alpha)$ is
determined by the convention used in choosing the sign of $Z_6$.
If $H^0$ is the SM-like Higgs boson, then \eq{d5} yields the condition for the decoupling limit,
\beqa
\sin(\beta-\alpha) &\simeq  & \frac{|Z_6| v^2}{m_H^2-m_h^2}\ll1\,,\\
m_h^2-m_A^2\ & \simeq & Z_5 v^2\,,
\eeqa
which can only be achieved for weak-coupling decoupling ($|Z_6|\ll 1$).
For completeness, we note that in the real Higgs basis,
$Z_5$ and $Z_6$ can be expressed in terms of the parameters of the scalar potential given in \eq{genpot},
\beqa
Z_5&=&\lambda_5+\tfrac{1}{4}(\lambda_1+\lambda_2-2\lambda_{345})\sin^2 2\beta\,,\\
Z_6&=&\pm\half\sin 2\beta(\lambda_1\cos^2\beta-\lambda_2\sin^2\beta-\lambda_{345}\cos 2\beta)\,,
\eeqa
where $\lambda_{345}\equiv \lambda_3+\lambda_4+\lambda_5$ and the sign ambiguity in $Z_6$ reflects the two-fold ambiguity of the real Higgs basis.

The MSSM Higgs sector at tree-level is a Type-II 2HDM.  The corresponding scalar potential is a special case of
\eq{genpot} with
\beq
\lambda_1=\lambda_2=-\lambda_{345}=\tfrac{1}{4}(g^2+g^{\prime\,2})\,,\qquad\quad
\lambda_4=-\half g^2\,,\qquad\quad \lambda_5=\lambda_6=\lambda_7=0\,.
\eeq
In terms of the scalar potential parameters in the real Higgs basis~\cite{Haber:2006ue},
\beqa
Z_1&=&Z_2=\tfrac{1}{4}(g^2+g^{\prime\,2})\cos^2 2\beta\,,\qquad\quad Z_3=Z_5+\tfrac{1}{4}(g^2-g^{\prime\,2})\,,\qquad\qquad
Z_4=Z_5-\half g^2\,,\nonumber \\
Z_5&=&\tfrac{1}{4}(g^2+g^{\prime\,2})\sin^2 2\beta\,,\qquad\qquad\quad\,\,\,\,
Z_6=-Z_7=\pm\tfrac{1}{4}(g^2+g^{\prime\,2})\sin 2\beta\cos 2\beta\,.\label{zsusy}
\eeqa
Note that the tree-level MSSM Higgs sector possesses a large-mass decoupling limit when $m_A\gg m_Z$ in which case
$h^0$ is a SM-like Higgs boson.   The weak-coupling decoupling limit can be achieved
at tree-level when $\sin 4\beta\simeq 0$ [cf.~\eq{zsusy}].  However, in this case the radiative corrections to
$Z_6$ cannot be neglected.  On the other hand, it is possible to find regions of MSSM parameter space where the
radiative corrections conspire to (approximately) cancel the tree level value of $Z_6$, in which case one of the
neutral CP-even Higgs bosons of the MSSM is SM-like while the other Higgs bosons need not be particularly heavy~\cite{Carena:2001bg}.

If $h^0$ is a SM-like Higgs boson, then its couplings can be treated to first order in the parameter that measures
the departure from the decoupling limit.  For the Type-I and II 2HDMs, this parameter is $\cos(\beta-\alpha)$.
We summarize the SM-like Higgs couplings normalized to the corresponding SM results in Table~\ref{tabdecouplings}.
\begin{table}[ht!]
\caption{Couplings of the SM-like Higgs boson $h^0$ of the CP-conserving 2HDM, normalized to
those of the SM Higgs boson, in the
decoupling limit.  The $HVV$ couplings apply to both
$VV=W^+ W^-$ or $ZZ$.
\label{tabdecouplings} \\}
\begin{tabular}{|c||c|c|}\hline
Higgs interaction & 2HDM coupling & \phantom{xxxxx} decoupling limit\phantom{xxxxx}\\
\hline
$hVV$ & $\sinbma$ & $1-\half\cosbmasq$
 \\[6pt]
$hhh$ & see eq.~(61) of Ref.~\cite{Haber:2006ue} &  $1+2(Z_6/Z_1)\cosbma$
  \\[6pt]
$hhhh$ & see eq.~(62) of Ref.~\cite{Haber:2006ue}  &  $1+3(Z_6/Z_1)\cosbma$
  \\[6pt]
$h\overline{D}D$~[\text{Type-I}]\,,\,$h\overline{U}U$~[\text{Types-I and II}]\, & $ \phantom{xxxxx}\phm\displaystyle\frac{\cos\alpha}{\sin\beta}=\sinbma+\cosbma\cot\beta$ \phantom{xxxxx} &
$1+\cosbma\cot\beta$
\\[6pt]
$h\overline{D}D$~[\text{Type-II}]\, &  \phantom{xxxxx}$-\displaystyle\frac{\sin\alpha}{\cos\beta}=\sinbma-\cosbma\tan\beta$ \phantom{xxxxx} &
$1-\cosbma\tan\beta$
\\[6pt] \hline
\end{tabular}
\end{table}

For example, in the Type-II 2HDM, if $\lambda_V$ is a Higgs coupling to vector bosons, $\lambda_H$ is the trilinear Higgs self-coupling,
and $\lambda_t$ [$\lambda_b$] are Higgs couplings to up-type
[down-type] fermions, then the couplings of $h^0$ approach the decoupling limit as shown below:
\beqa
\frac{\lambda_{V}}{[\lambda_{V}]_{\rm SM}}&=&1+\mathcal{O}\left(\frac{Z_6^2m_Z^4}{(m_H^2-m_h^2)^2}\right)\,,\\
\frac{\lambda_{H}}{[\lambda_{H}]_{\rm SM}}&=&1+\mathcal{O}\left(\frac{Z_6^2 m_Z^2}{m_H^2-m_h^2}\right)\,,\\
\frac{\lambda_{t}}{[\lambda_{t}]_{\rm
    SM}}&=&1+\mathcal{O}\left(\frac{Z_6 m_Z^2\cot\beta}{m_H^2-m_h^2}\right)\,.\\
\frac{\lambda_{b}}{[\lambda_{b}]_{\rm
    SM}}&=&1+\mathcal{O}\left(\frac{Z_6 m_Z^2\tan\beta}{m_H^2-m_h^2}\right)\,.\label{uf}
\eeqa
That is, the approach to decoupling is fastest in the case of the
$h^0 VV$ couplings and slowest in the case of the Type-II $h^0 bb$ couplings at
large $\tan\beta$.  Moreover, the approaches to large-mass decoupling and weak-coupling decoupling
are only distinguished by the behavior of the Higgs self-coupling.  Finally, we note that
if only $h^0$ is light (implying the large-mass decoupling limit), then
a precision measurement of the
$h^0b\bar{b}$ coupling provides the greatest sensitivity to the mass scale of the heavy Higgs states~\cite{Carena:2001bg}.

\section{Near-mass-degenerate scenarios}

The LHC Higgs data suggest that the Higgs boson has SM-like couplings to vector boson pairs.  In light of this result, how would one
interpret future deviations (if present) from SM-like Higgs couplings in fermionic decay channels or the $\gamma\gamma$ decay channel in a 2HDM framework?
The results of of Table I suggest that deviations in the Higgs couplings to down-type fermions from SM predictions
could be a result of first order non-decoupling effects that can be significant in Type-II models at large $\tan\beta$.
A deviation in the effective Higgs couplings to $\gamma\gamma$ could be due to the contributions of new charged states
that appear in the loop~\cite{manyrefs1}.  In this section, we wish to explore a third possibility in which more than one neutral Higgs state
is responsible for the Higgs data observed at the LHC~\cite{degenrefs1,Ferreira:2012nv,degenrefs2,degenrefs3}.

If there are two (or more) nearly mass-degenerate neutral Higgs
states, then the large-mass decoupling limit is not relevant.
Consider as an example the case of a near mass-degenerate $h^0$, $A^0$
pair in the Type-I or Type-II 2HDM.  In this case, the coupling of
$h^0$ to vector boson pairs can be approximately SM-like in the
weak-coupling decoupling limit.  Moreover, in this case it is possible
to have near-mass-degenerate neutral Higgs states.  We shall assume
that the two states are close enough (within about 1 GeV) in mass so
that the two states cannot be resolved with the present Higgs data.  The
$ZZ^*\to 4$~leptons and the $WW^*$ final states would originate from
the decay of $h^0$ (which is assumed to be SM-like), since the
tree-level couplings of $A^0$ to vector boson pairs are absent.
Similar scenarios also exit for a near-mass-degenerate $H^0$, $A^0$
pair (where $H^0$ is the SM-like Higgs boson) and for a
near-mass-degenerate $h^0$, $H^0$ pair.  In the latter case, if $h^0$
is the SM-like Higgs boson, then the unitarity sum rule would imply
that the couplings of $H^0$ to vector boson pairs are
suppressed~\cite{Gunion:1990kf} .
Indeed in the case of a a near-mass-degenerate $h^0$, $H^0$ pair, the
unitarity sum rule implies that the two states taken together would
exhibit couplings to vector boson pairs that are indistinguishable
from those of the SM-like Higgs boson (assuming that the individual neutral
Higgs states cannot be separately resolved).

The phenomenology of near-mass-degenerate Higgs states has been treated in Refs.~\cite{degenrefs1,Ferreira:2012nv,degenrefs2}.
Here, I shall outline the results obtained in Ref.~\cite{Ferreira:2012nv}.  We performed parameter scans of the Type-I and II
2HDMs under the assumptions that:
(i) two of the neutral Higgs states are nearly degenerate in mass, with a common mass taken
to be the mass of the observed Higgs boson, $m_{h^0}\simeq 125~{\rm GeV}$;
(ii) the scalar potential is bounded from below~\cite{Deshpande:1977rw};
(iii) the couplings of $h^0$ to $VV$ ($V=W$ or $Z$) are within 20\% of their SM-values; and
(iv) precision electroweak constraints on the 2HDM contributions to the Peskin--Takeuchi~\cite{Peskin:1990zt} 
$S$, $T$ and $U$ parameters
are satisfied~\cite{Froggatt:1991qw,Haber:2010bw}.
These assumptions are further
subjected to the constraints imposed by the following experimental observables:
$b\to s\gamma$,
$B_d^0$--$\overline{B}_d^0$ and $B_s^0$--$\overline{B}_s^0$ mixing,
$R_b\equiv \Gamma(Z\to b\bar{b})/\Gamma(Z\to {\rm hadrons})$ and
$B^+\to\tau^+\nu_\tau$.


\begin{figure}[t!]
\includegraphics[width=0.49\textwidth]{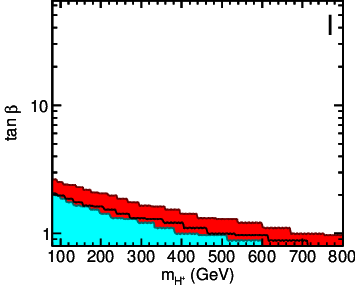}
\includegraphics[width=0.49\textwidth]{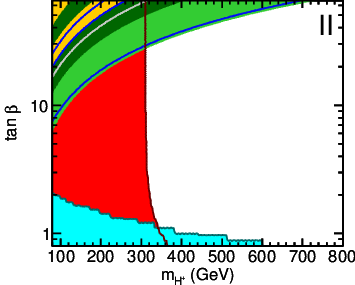}
\caption{
Excluded regions of the ($m_{H^+}\,,\,\tan\beta$) parameter space for the Type I and II 2HDMs.
The color coding is as follows: $\mathrm{BR}(B\to X_s\gamma)$ (red), $\Delta_{0-}$ (black contour),
$\Delta M_{B_d}$ (cyan), $B_u\to\tau\nu_\tau$ (blue), $B\to D\tau\nu_\tau$ (yellow),
$K\to\mu\nu_\mu$ (gray contour), $D_s\to\tau\nu_\tau$ (light green),
and $D_s\to\mu\nu_\mu$ (dark green).  Taken from Ref.~\cite{Mahmoudi:2009zx}.
\label{fig:types}
}
\end{figure}

Some of the important constraints on the parameter space for Type I and II 2HDMs
are summarized in Fig.~\ref{fig:types}, where excluded parameter regions derived from
$\mathrm{BR}(B\to X_s\gamma)$, $\Delta_{0-}$, $\Delta M_{B_d}$, $B_u\to\tau\nu_\tau$,
$B\to D\tau\nu_\tau$, $K\to\mu\nu_\mu$, $D_s\to\tau\nu_\tau$, and $D_s\to\mu\nu_\mu$
are plotted in the ($m_{H^+}$\,,\,$\tan\beta$) plane~\cite{Mahmoudi:2009zx}.
Here, we have included among the list of flavor observables the degree
of isospin asymmetry in the exclusive decay mode $B\to K^*\gamma$,
defined as~\cite{Kagan:2001zk}
\begin{equation}
\Delta_{0-}\equiv\frac{\Gamma(\overline{B}\llsup{0}\to\overline{K}\llsup{*0})-
\Gamma(\overline{B}\llsup{\,-}\to\overline{K}\llsup{*\,-})}
{\Gamma(\overline{B}\llsup{0}\to\overline{K}\llsup{*0})-
\Gamma(\overline{B}\llsup{\,-}\to\overline{K}\llsup{*\,-})}\,.
\end{equation}
The exclusion of low $\tan\beta< 1$ in the Type I and II 2HDMs for $m_{H^+}<500$~GeV,
arises as a result of three observables: $\mathrm{BR}(B\to X_s\gamma)$, $\Delta_{0-}$, and $\Delta M_{B_d}$.
In the Type-I 2HDM,
a value of $\tan\beta>1$ signals the decoupling of one Higgs doublet from the whole fermion sector.
In the Type-II 2HDM, there
exists a $\tan\beta$-independent lower limit of $m_{H^+}\gtrsim 300$ GeV imposed by
$\mathrm{BR}(B\to X_s\gamma)$.  (This latter constraint is now
somewhat more stringent in light of Ref.~\cite{Hermann:2012fc}.)
No generic lower limit on $m_{H^+}$ is found in the Type I 2HDM.
Constraints for high $\tan\beta$ are only obtained in the Type II 2HDM.
This can be understood by noting that the leptonic and semi-leptonic observables
require $\tan\beta$-enhanced couplings to down-type fermions.

Finally, recently current data from B\textsc{a}B\textsc{ar} of the $\bar B \to D \tau
\bar \nu$ and $\bar B \to D^\ast \tau \bar \nu$ slightly deviate from
the Standard Model predictions by 2.0 $\sigma$ and 2.7 $\sigma$,
respectively~\cite{Lees:2012xj}.  Moreover, these data are also inconsistent with the
Type-I and Type-II 2HDMs, since both decay rates, which depend
on the charged Higgs mass,  cannot be
explained simultaneously for the same value of $m_{H^+}$.
However, there is no
confirmation yet of the BaBar results for $\bar B \to D \tau \bar\nu$
and $\bar B \to D^\ast \tau \bar \nu$ from the BELLE collaboration.
In the present analysis, these data will not be included in our constraints.

We now consider the observables that can be extracted from the Higgs data.  It is convenient to define
$$
R_f^H=\frac{\sigma(pp\to H)_{\rm 2HDM}\times {\rm BR}(H\to f)_{\rm 2HDM}}
{\sigma(pp\to h_{\rm SM})\times {\rm BR}(h_{\rm SM}\to f)}\,,
$$
where $f$ is the final state of interest, and $H$ is one of the two 125 GeV mass-degenerate
scalars.  The observed ratio of $f$ production relative to the SM expectation is
$$
R_f\equiv\sum_H R_f^H\,.
$$
Two main Higgs production mechanisms, $gg$ fusion and vector boson
($W^+W^-$ and $ZZ$) fusion, are included in our analysis.  The final states of interest are $f=\gamma\gamma$, $ZZ^*$, $WW^*$ and $\tau^+\tau^-$.
Note that the LHC is (eventually) sensitive to the $b\bar{b}$ final state primarily in associated $Vh$ production
(where $V=W^\pm$ or $Z$), which will be briefly considered below.
\begin{figure}[t!]
\includegraphics*[width=0.49\textwidth]{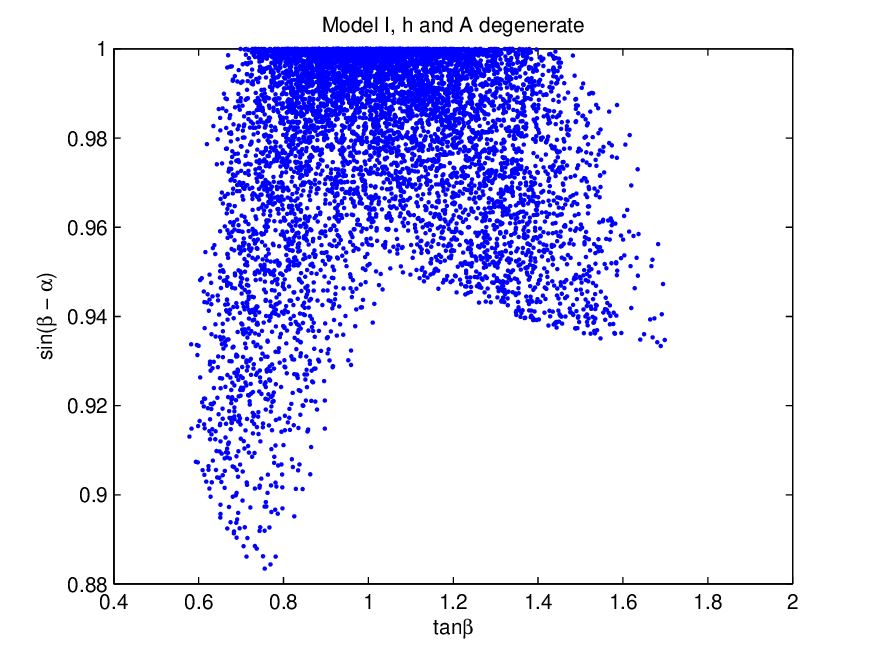}
\includegraphics*[width=0.49\textwidth]{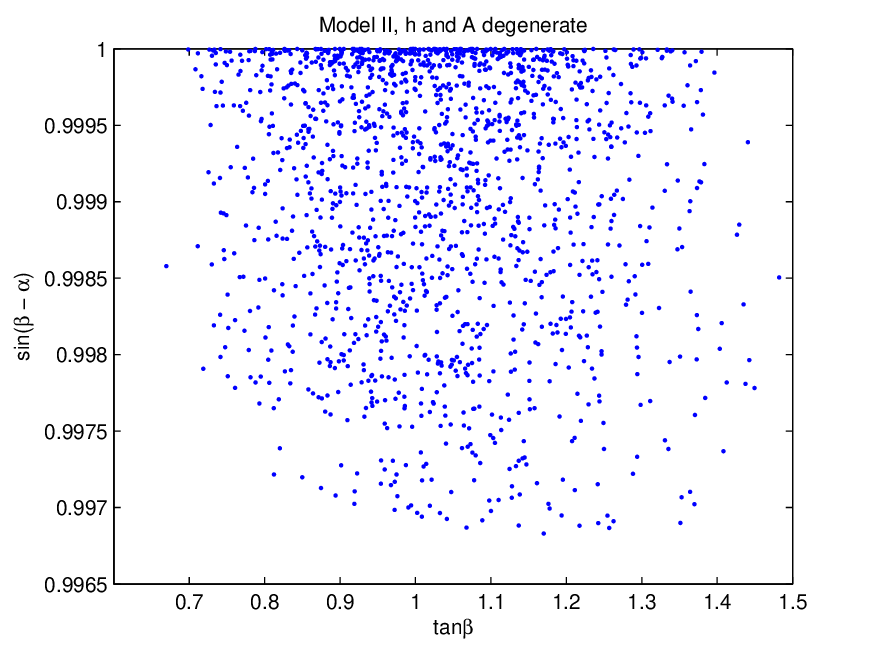}
\caption{\label{sbma}
Values of $\sin(\beta-\alpha)$ as a function of $\tan\beta$ in the 2HDM of
Type-I (left panel) and Type-II (right panel), subject to $0.8<R_{ZZ}<1.2$ and
the near-mass degeneracy of $h^0$ and $A^0$.  The value of $\sin(\beta-\alpha)=1$
corresponds to limit in which $h^0$ possesses SM couplings to gauge bosons and fermions.
}
\end{figure}
In our analysis, we assume that $R_{WW}\simeq R_{ZZ}\simeq 1\pm 0.2$.

We first consider the implications of an
enhanced $\gamma\gamma$ signal due to mass-degenerate $h^0$ and $A^0$.
By imposing the constraints of the mass-degenerate $h^0$, $A^0$ pair, we find that
$\sin(\beta-\alpha)$ is necessarily near 1 as shown in Fig.~\ref{sbma}. Hence, it follows that the couplings of $h^0$
to the massive gauge boson pairs are close to their SM values. Similar result
follow for other mass-degenerate pair choices.

\begin{figure}[t!]
\begin{center}
\includegraphics*[width=0.47\textwidth]{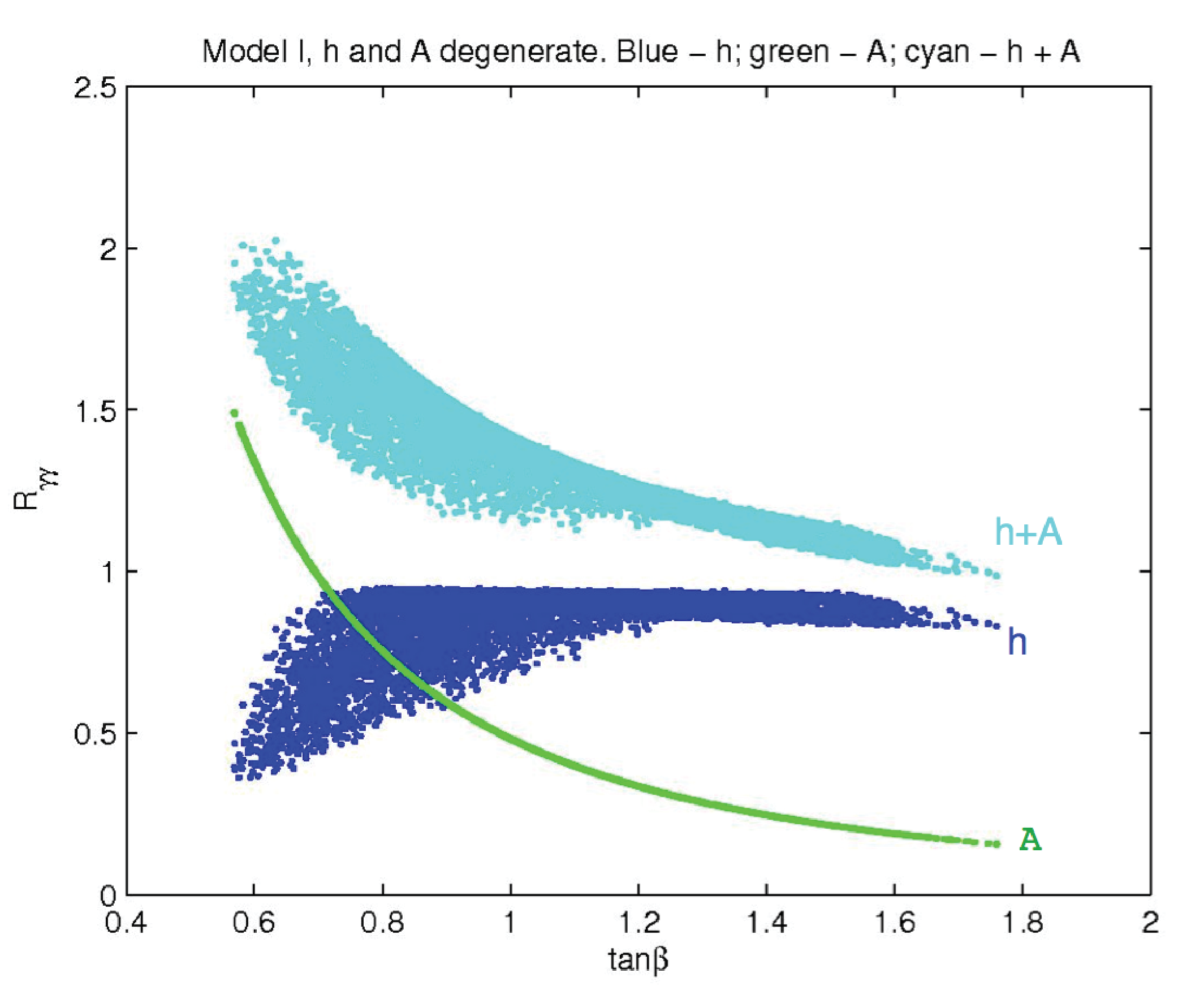}
\includegraphics*[width=0.51\textwidth]{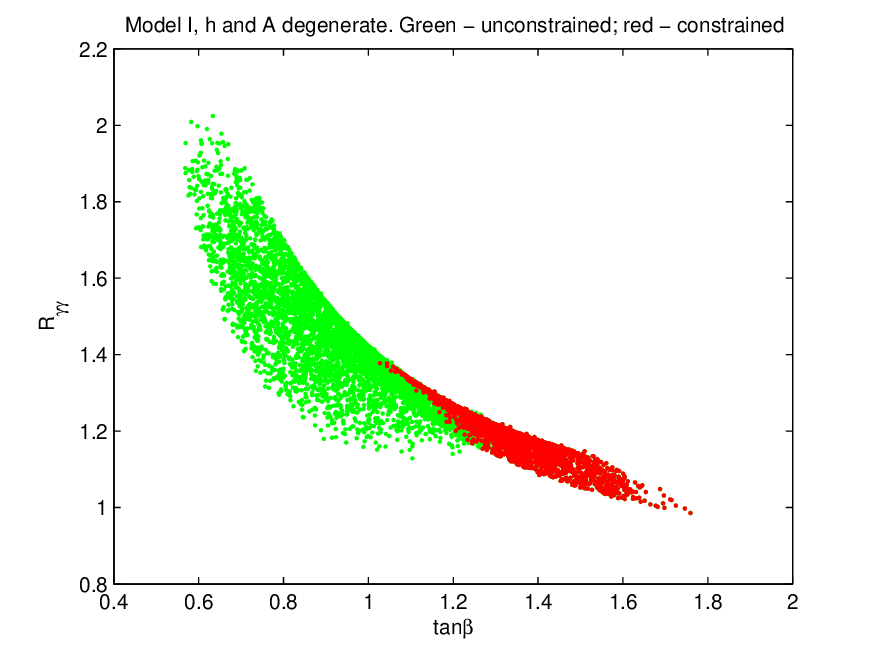}
\end{center}
\caption{For the Type-I 2HDM.  Left panel: $R_{\gamma \gamma}$ as a function of $\tan{\beta}$ for
$h$ (blue),
$A$ (green),
and the total observable rate (cyan),
obtained by summing the rates with intermediate $h$ and $A$,
for the unconstrained scenario.
Right panel:  Total rate for $R_{\gamma \gamma}$ as a function of $\tan{\beta}$
with (red) and without (green) the $B$-physics constraints..
\label{fig:hA_modI_phph}}
\end{figure}

\begin{figure}[t!]
\begin{center}
\includegraphics*[width=0.51\textwidth]{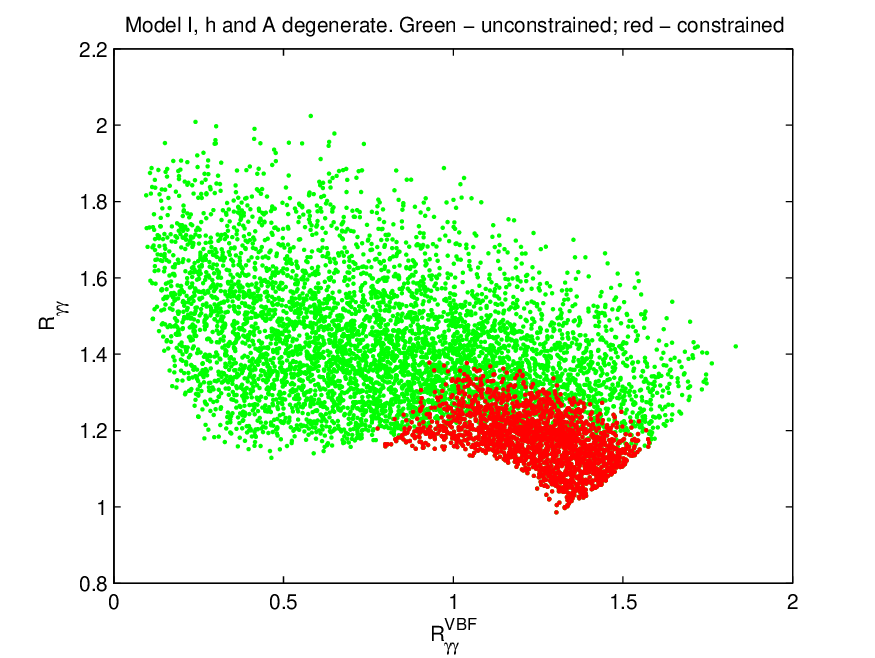}
\includegraphics*[width=0.48\textwidth]{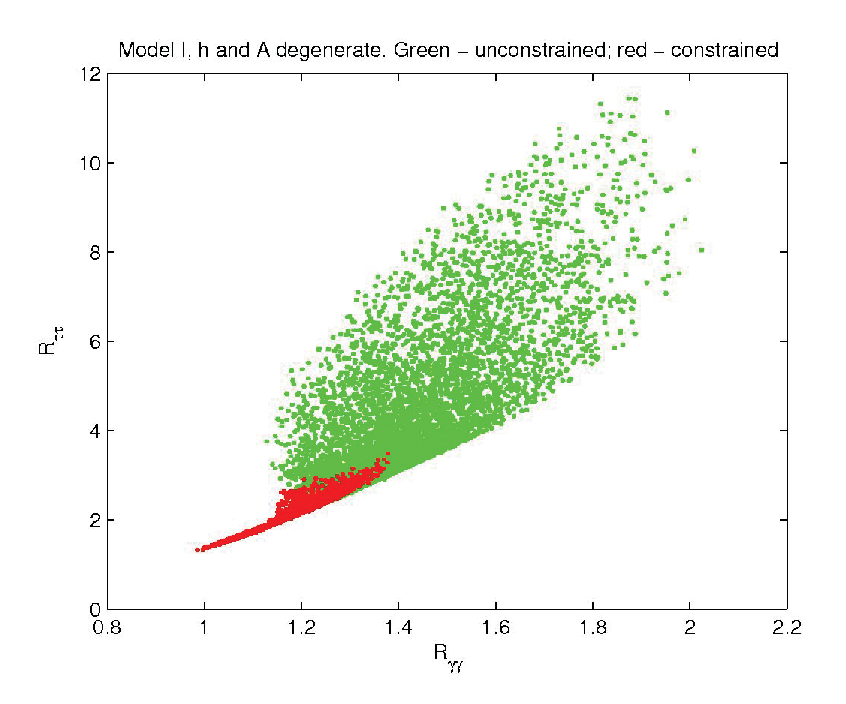}
\end{center}
\caption{For the Type-I 2HDM.   Left panel: allowed region in the
$R^{\textrm{VBF}}_{\gamma \gamma} - R_{\gamma \gamma}$
plane with (red) and without (green) the $B$-physics constraints.
Right panel: total $R_{\tau\tau}$ ($h$ and $A$ summed)
as a function of $R_{\gamma \gamma}$
for the constrained (red) and unconstrained (green) scenarios.
\label{typeonedegen}}
\end{figure}

Consider first the case of the Type-I 2HDM.  
The enhancement of $R_{\gamma\gamma}$ occurs in the parameter regime of $\tan\beta\lsim 1.5$,
as shown in Fig.~\ref{fig:hA_modI_phph}.
It is also possible to experimentally separate out $\gamma\gamma$ events that arise from Higgs
bosons produced by $WW$-fusion.  The relevant quantity of interest is
\beq
R_{\gamma\gamma}^{\rm VBF}=\frac{\sigma(pp\to VV\to h)_{\rm 2HDM}\,{\rm BR}(h\to
\gamma\gamma)_{\rm 2HDM}}{\sigma(pp\to VV\to h_{\rm SM})\,{\rm BR}(h_{\rm SM}\to
\gamma\gamma)}\,.
\eeq
There is some correlation between $R_{\gamma\gamma}$ and $R_{\gamma\gamma}^{\rm VBF}$ as
shown in the left panel of Fig.~\ref{typeonedegen} in which both quantities exceed unity.  Note that there 
are also allowed parameter points in which only one of these two quantities is enhanced above
SM expectations.

An enhanced $\gamma\gamma$ signal in the Type-I 2HDM for the mass-degenerate scenario yields
a number of associated predictions that must be confirmed by experiment if this
framework is to be consistent.  First, the inclusive $\tau^+\tau^-$ signal is enhanced with respect to the
SM due to the production of $A$ via $gg$ fusion, as shown in the right panel of Fig.~\ref{typeonedegen}.
A similar enhancement would occur in the inclusive $b\bar{b}$ production from Higgs decay, although this channel
cannot be isolated at the LHC due to huge QCD backgrounds.  In contrast,
the exclusive $b\bar{b}$ signal due to the production of Higgs bosons in association with $V=W^\pm$ or $Z$ is close to its
SM value but is not enhanced, due to the assumed SM-like $hVV$ coupling.  

\begin{figure}[t!]
\begin{center}
\includegraphics*[width=0.52\textwidth]{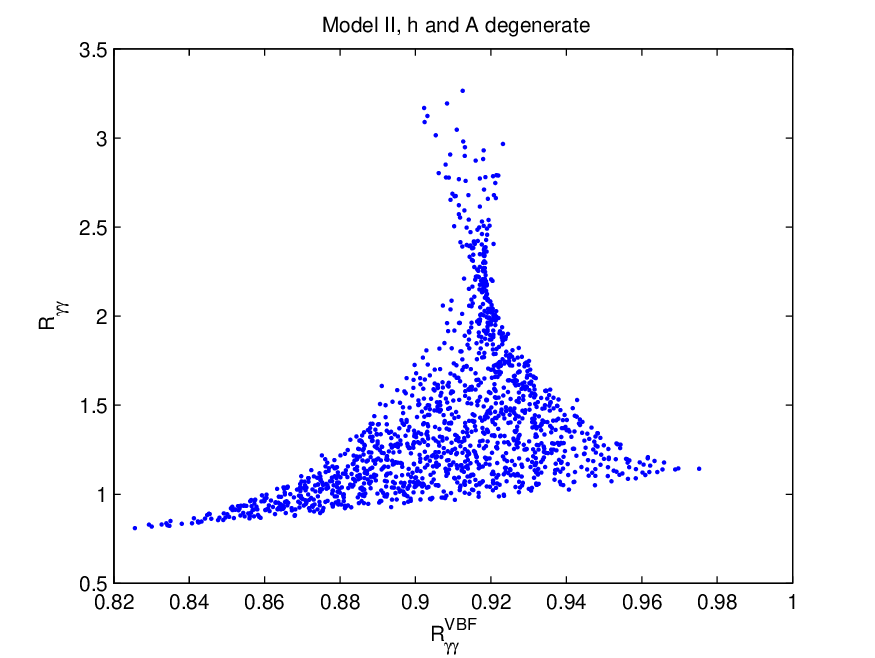}
\includegraphics*[width=0.47\textwidth]{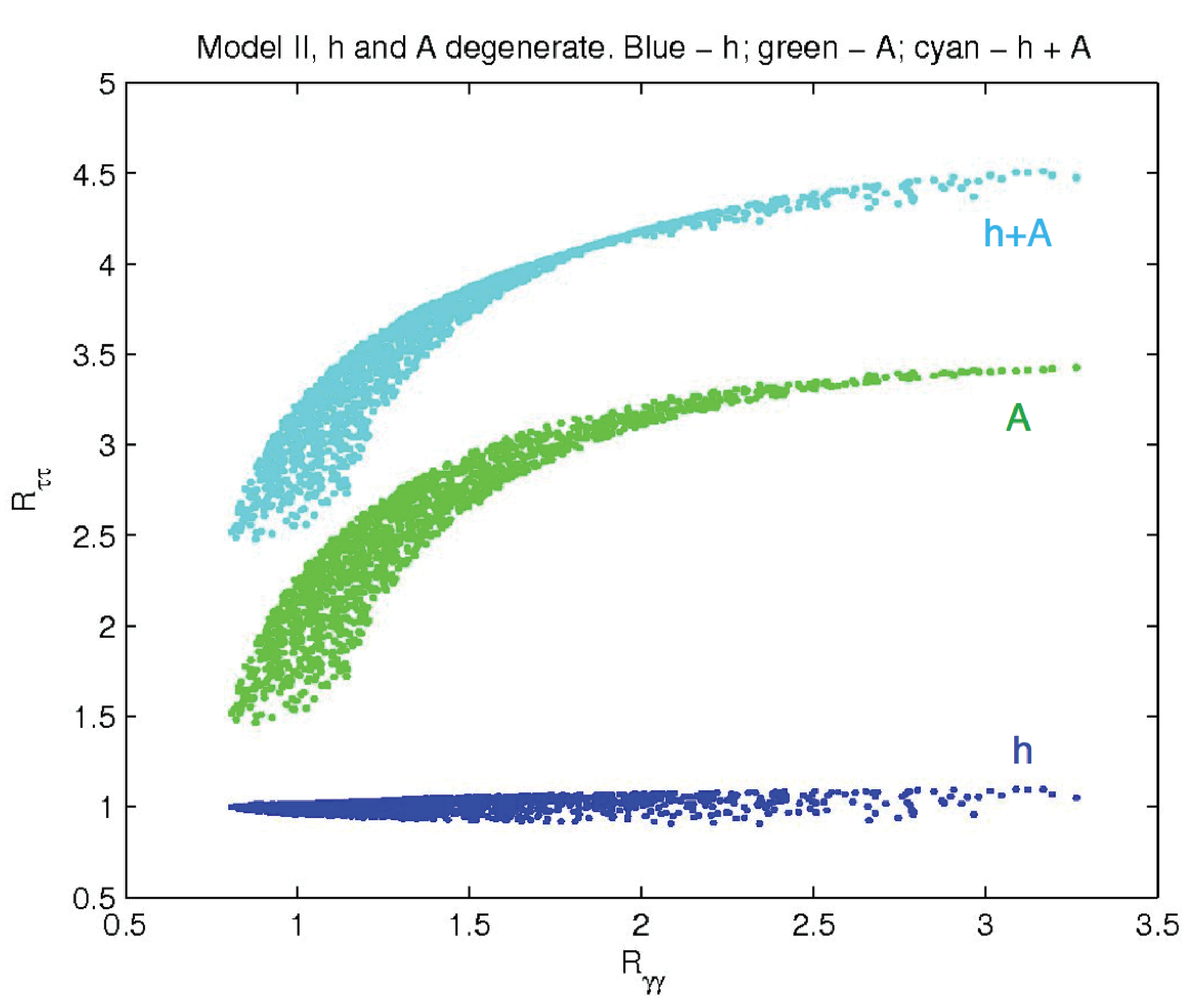}
\caption{For the Type-II 2HDM.  Left panel: Allowed region in the
$R^{\textrm{VBF}}_{\gamma \gamma} - R_{\gamma \gamma}$.
Right panel: $R_{\tau \tau}$ as a function of $R_{\gamma \gamma}$ for
$h$ (blue),
$A$ (green),
and the total observable rate,
obtained by summing the rates with intermediate $h$ and $A$ (cyan).
\label{fig:vbf_vs_totII}\\[-20pt]}
\end{center}
\end{figure}

We can repeat the exercise for the Type-II 2HDM. Once we assume a heavy
charged Higgs mass, there are no further constraints from $B$ physics.
In this case, $R_{\gamma\gamma}^{\rm VBF}$
can never be enhanced above 1 as shown in the left panel of Fig.~\ref{fig:vbf_vs_totII}, since it only
receives contributions from $h$ production, which has nearly exact SM couplings
since $\sin(\beta-\alpha)$ is extremely close to 1 [cf.~the right panel of Fig.~\ref{sbma}].
As in the Type-I 2HDM, the $\tau^+\tau^-$ signal is also enhanced as shown in the right panel of Fig.~\ref{fig:vbf_vs_totII}, which is a critical prediction
of the mass-degenerate scenario.

\begin{figure}[t!]
\includegraphics[width=0.49\textwidth]{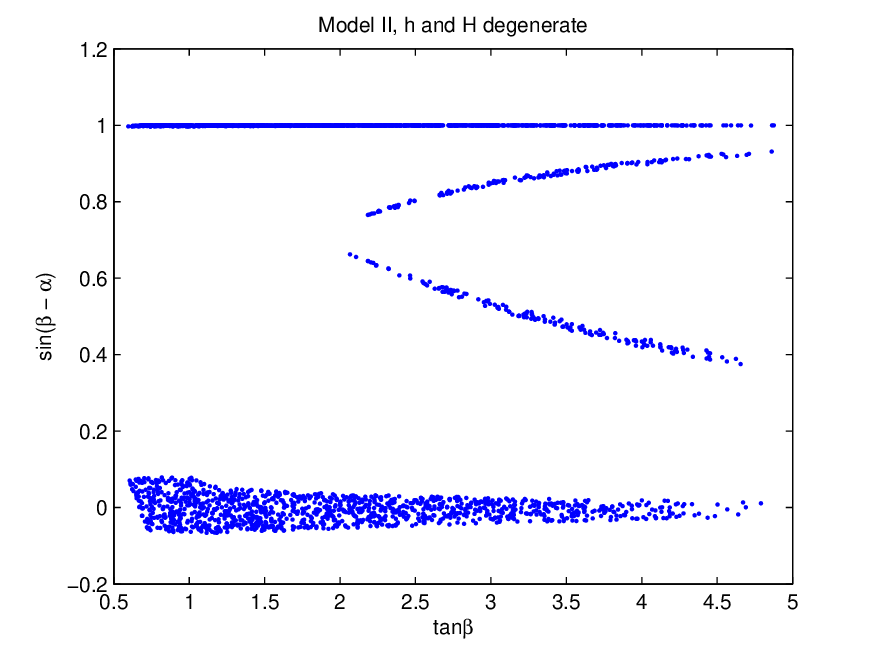}
\includegraphics[width=0.49\textwidth]{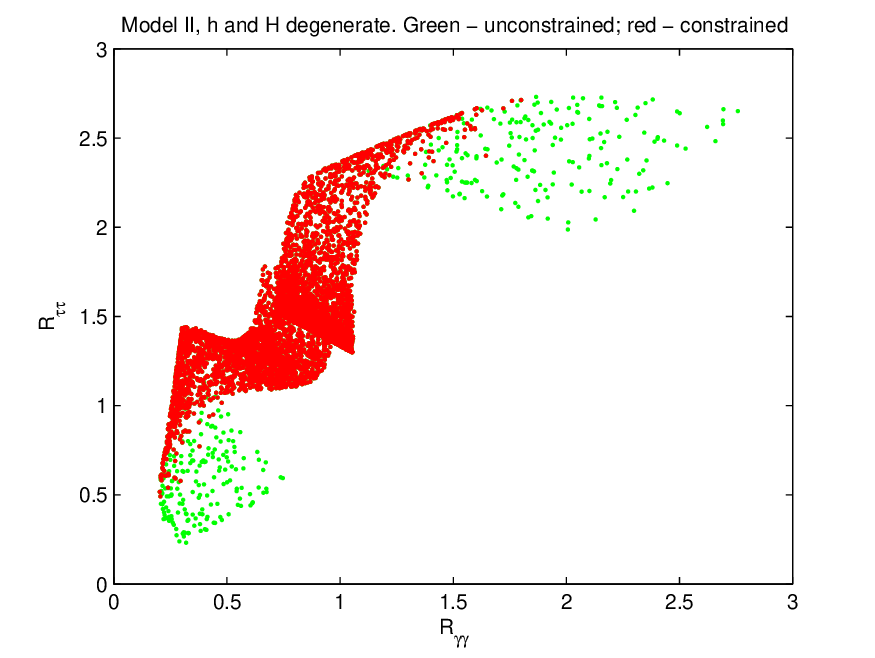}
\caption{Left panel: Values obtained in the $\tan\beta$---$\sin(\beta-\alpha)$ plane for the generated points that satisfy $0.8 < R_{\gamma\gamma} < 1.5$.
Right panel: Values for $R_{\gamma\gamma}$ as a function of $\tan\beta$ with (red) and without (green) the $B$-physics constraints.
\label{fig7}}
\end{figure}

For completeness, the implications of other nearly-mass-degenerate neutral Higgs pairs are briefly summarized here (details can be found in Ref.~\cite{Ferreira:2012nv}).  Consider the
case of a nearly mass-degenerate $h^0$ and $H^0$.  In the Type-I 2HDM, there is no longer a constraint on $\sin(\beta-\alpha)$ since both
$h$ and $H$ can couple to vector boson pairs. It turns out that after imposing
$B$-physics constraints, it is not possible to enhance the $\gamma\gamma$ signal.
In Type-II models, the constraint on $\sin(\beta-\alpha)$ is more complicated as shown in the left panel of Fig.~\ref{fig7}. We find
that an enhanced $\gamma\gamma$ signal is possible.
As in the previous case of mass-degenerate scalars in the Type-II 2HDM, the $\gamma\gamma$
signal resulting from Higgs bosons produced in vector boson fusion is slightly
suppressed.   Likewise, the $\tau^+\tau^-$ signal is again enhanced in regions of the enhanced
$\gamma\gamma$ signal, as shown in the right panel of Fig.~\ref{fig7}.

The case in which $H^0$ and $A^0$ constitute the nearly-mass degenerate pair (with $H^0$
responsible for the observed Higgs decay to $ZZ^*$ and $WW^*$)
would imply that a lighter Higgs state $h^0$ was missed at LEP, which is
possible if the $h^0 ZZ$ coupling is sufficiently suppressed.  We assume that $m_{H^0}<2m_{h^0}$;
otherwise $H^0\to h^0 h^0$ would be a significant decay mode and the $H^0\to ZZ^*\to 4$~leptons
signal would be suppressed.    Assuming that $m_{H^0}\simeq m_{A^0}\simeq$~125~GeV and $m_{h^0}>\half m_{H^0}$,
the LEP Higgs search cannot probe regions of $\sin(\beta-\alpha)\lsim 0.1$~\cite{Barate:2003sz}.
In the case of the mass-degenerate $H^0$ and $A^0$, we find that $m_{H^+}$ must lie below about 200 GeV; otherwise, the Higgs corrections to the electroweak 
$\rho$-parameter are too large~\cite{Grimus:2007if}.  This immediately rules out the Type-II 2HDM (due to $b\to s\gamma$ constraints), so we have examined the consequences of a  nearly-mass degenerate $H^0$ and $A^0$ pair for the Type-I 2HDM.
As before, the enhanced $\gamma\gamma$ signal lies mostly in the vicinity of $\tan\beta\lsim 1$
(where the $\tau^+\tau^-$ signal is also enhanced).  However, taking all $B$-physics constraints into account, only a few points with a (slightly) enhanced $\gamma\gamma$ signal survive.

Ultimately, if a near-mass degenerate neutral Higgs pair exists with a common mass around 125 GeV, it would be surprising if the mass difference were significantly smaller than a GeV.  In this case, one would expect that evidence for a second state could be revealed with a large enough data sample.
Current Higgs mass measurements in the $ZZ^*\to 4$~leptons and $\gamma\gamma$ channels
have mass resolutions in the range of $1$--$2$~GeV.  In models of an enhanced $\gamma\gamma$ signal due to nearly-mass-degenerate states, the
$ZZ^*\to 4$ leptons channel arises entirely from one SM-like Higgs boson state,
whereas the $\gamma\gamma$ signal is made up of contributions from both scalar states.
Thus, the average mass inferred from the $\gamma\gamma$ channel can be slightly different
from the one inferred from the $ZZ^*\to 4$ leptons channel.
Indeed, the present Higgs data sample does seem to indicate slightly different masses
in the $ZZ^*\to 4$~leptons and $\gamma\gamma$ channels (albeit not at a statistically significant level with the present data), although the sign of the mass difference differs in the ATLAS and CMS data samples.

\section{Conclusions}

The current LHC Higgs data sets are limited in statistics.  Despite some intriguing
variations, the present data is consistent with a SM-like Higgs boson.
If further data reveal no statistically significant deviations from SM Higgs behavior,
then we are in the domain of the decoupling limit. A precision Higgs program is then
required to elucidate the possibility of new Higgs physics beyond the Standard Model~\cite{Asner:2013psa}.

In the two-Higgs-doublet model (2HDM), one expects the approach to the decoupling limit to be fastest
in the behavior of the Higgs coupling to $W^+W^-$ and $ZZ$.  Present Higgs data already suggest that these couplings are within about $20$--$30\%$ of SM expectations.   Even if no significant deviations in these couplings are detected in future LHC data, it is still possible that deviations from SM predictions could emerge in the Higgs couplings to $\gamma\gamma$ and down-type fermion pairs.  The enhancement in the $\gamma\gamma$ channel observed by the ATLAS Collaboration has fueled much speculation of such a possibility, which would be a clear hint of new physics beyond the Standard Model if confirmed.

One possible source of an enhancement (if present)
in the $\gamma\gamma$ channel of the Higgs signal (while maintaining SM-like Higgs couplings to $W^+W^-$ and $ZZ$)
is the existence of a mass-degenerate pair of neutral Higgs bosons in the 2HDM~\cite{Ferreira:2012nv}.  We find that a significant
enhancement can occur in the 2HDM with Type-I and Type-II Higgs-fermion Yukawa couplings for values
of $\tan\beta$ near 1 (or below).
Such a scenario is easily tested, as it would also require
an enhanced production of $\tau^+\tau^-$ and the
possibility of a small measurable difference in the Higgs mass measurement in the
$ZZ^*\to 4$~leptons and $\gamma\gamma$ channels.



\begin{acknowledgments}

I gratefully acknowledge my collaborators Pedro Ferreira, Rui Santos and Joao Silva
and their contributions to the work on nearly-mass-degenerate neutral Higgs states
that was presented in the second half of this talk.
This work was supported in part by U.S. Department of
Energy grant number DE-FG02-04ER41286.
\end{acknowledgments}

\end{document}